\documentclass[noshowpacs, twocolumn,preprintnumbers,secnumarabic,amsmath,amssymb,nofootinbib]{revtex4-1}
\usepackage{dcolumn}      
\usepackage{bm}           
\usepackage{graphicx}
\usepackage{natbib}
\usepackage{caption}
\usepackage{wrapfig}
\usepackage{float}
\usepackage{subcaption}
\usepackage{amsmath,amssymb}  
\usepackage{bm}  
\usepackage[font=small,labelfont=bf]{caption}

\usepackage[pdftex,bookmarks,colorlinks,breaklinks]{hyperref}  
\usepackage{mathrsfs}
\usepackage{amsmath}

\begin{document}
\title{Euler potentials and Magnetic fields}
\author{Bob Osano}
\email{bob.osano@uct.ac.za} \affiliation{Cosmology and Gravity Group, Department of Mathematics and Applied Mathematics, University of Cape Town, Rondebosch 7701, Cape Town, South Africa\\\\}
\affiliation{Academic Development Programme, Science, Centre for Higher Education Development, University of Cape Town, Rondebosch 7701, Cape Town, South Africa}

\begin{abstract}
It is known that the cross product of gradients of two scalars, the Euler potentials (EP), may represent magnetic fields lines. We examine the utility of such potential in the broader magneto-genesis and dynamo theories, and find that a reinterpretation of the potentials offer a new understanding of the role EP may play in the evolution of magnetic fields.
\end{abstract}
\date{\today}

\maketitle
\section{Introduction}
In every direction in space that we point out instruments, we find properties that could be ascribed to the possible presence of magnetic fields or influence thereof. These are seen via synchrotron emission, polarization or Faraday rotation at radio frequency of between $0.2$ and $10$ GHz \cite{Beck}. Studies of cosmic magnetic fields have a long history and are driven by the need to explain what is observed and the desire to know what role magnetic fields play in the formation and evolution of cosmic structures ranging from stars and galaxies to super clusters and the universe. Nevertheless, we still do not have a clear handle on how the magnetic fields arise or how they evolve. These two outstanding problems are often classified and studied under the topics (i) {\it magneto-genesis}, and (ii) amplification of magnetic fields mainly via a {\it dynamo} mechanism. 

It is thought that original magnetic fields, or seed fields as they are commonly called, may have their origin in phase transition  in the early universe \cite{caprini, widrow}, or in the cosmological structure formation \cite{lazar}, or in the first stars and blackholes \cite{Rees}, or  in the first supernova\cite{Hanayama} or even as result of fluctuations in the gravitational field \cite{Bishop}. Once generated these fields are thought to be sustained primarily by a dynamo action\cite{Beck, Axel}.

The subject of this brief article has to do with the dynamo effect. Research on dynamo action has a long history that goes back to the work of \cite{larmor} where a homogeneous dynamo mechanism was proposed as a possible source of magnetic fields observed in the sunspots. The proposal remained largely ignored particularly because of the results published soon after in \cite{Cow1} which  showed that it was impossible to generate axi - symmetric or two-dimensional magnetic fields via the homogeneous dynamo process. 
This became known as {\it Cowling's theorem}. It was not until after \cite{Back1} and \cite{Her} were published that the ideas in \cite{larmor} were salvaged. Whereas Cowling showed that it was not possible to generate axi-symmetric magnetic fields, the restriction did not apply to the non-axisymmetric ones. Backus \cite{Back1} and Herzenberg \cite{Her} demonstrated that non-axisymmetric fields could indeed be generated via a homogeneous dynamo. In effect, they presented arguments for the threshold necessary for the dynamo action. These, and other works on the threshold, are classified as {\it bounding theorems} as they provide lower bounds for this kind of dynamo. The questions then changed from whether or not a dynamo action could take place to what threshold was needed to induce the action.

The dynamo term in the magnetic induction equation is the term $\nabla\times( {\bf U}\times{\bf B})$, where ${\bf U}$ is the velocity and ${\bf B}$ is the magnetic flux. It is clear that the larger this term is, the greater the amplification of the magnetic field. Different forms of bounding theorems appear in literature giving special cases and designed for specific problems. In general, the velocity field has to be strong enough in order to stretch the magnetic fields to the point where they overcome Ohmic dissipation \cite{Back1,Child1,Proc1, By1, Proc2, Willis1, Chen1}. Most of these analysis are mathematical, but mathematical analysis alone without experimentation can only yield limited understand of the dynamo mechanism. On the other hand, the likelihood of replicating some of the extreme conditions needed for dynamos are slim. Computer simulations has emerged as a powerful alternative and one that is contributing to greater understanding dynamos. Simulations also offer a platform for examining magnetic analogues \cite{bob1, bob2, bob3, bob4} or testing predictions that result from any new formulation of electrodynamics, for example the formulation given in \cite{bob5} . In the remaining sections, we revisit the theoretical basis for the dynamo action and comment on Euler potentials in relation to the evolution of magnetic fields. Simulations related to Euler potentials have recently appeared \cite{Axelp, Dolag}.

\section{Euler potentials}
The Coulomb gauge is the cornerstone of Maxwell equations and we need to understand it in relation to vector potentials. In particular, the magnetic vector potential ${\bf A}$ (hereafter ${\bf A}_{\it MP}$) is chosen such that it obeys the Coulomb gauge i.e. $\nabla.{\bf A}_{\it MP}=0$, otherwise one can add a gradient of any scalar $\nabla\Xi$ to obtain a variation of the vector potential ${\bf A}'= {\bf A}_{\it M}+\nabla\Xi$. The problem here lies with the fact that $\nabla\times{\bf A}'=\nabla\times{\bf A}_{\it MP}={\bf B}$. Any spurious gauge functions such as $\Xi$ does not lead to a magnetic fields but has the consequence that $ \nabla. {\bf A}'\ne 0,$ other than for special cases as we shall demonstrate. So lets go back to the main point of this article which has to do with the dynamo equations in the presence of scalar functions. A good starting point is with Euler potentials \cite{Euler, Truesdell, Stern}. These scalar functions are conserved along field lines such that they allow for the description of magnetic field lines. The field lines can be thought of as given by the intersections of two surfaces defined by the gradients of the potentials. In particular, we define two scalar potential function $\alpha$ and $\beta$ with the properties $\nabla\alpha . {\bf B}=0=\nabla\beta . {\bf B}$. In effect, both gradient scalars are orthogonal to the field lines and in addition are orthogonal to each other such that 
\begin{eqnarray}\label{zetadef}\nabla\alpha\times\nabla\beta=\zeta {\bf B},\end{eqnarray}
where the coefficient $\zeta$ is a pure scalar quantity which may depend on time. Another way of looking at this is that we have a vector potential function ${\bf A}_{\it EP}=\alpha\nabla\beta,$ which allows the definition \begin{equation} \zeta{\bf B}=\nabla\times{\bf A}_{\it EP}=\nabla\times(\alpha\nabla\beta)=\nabla\alpha\times\nabla\beta.\end{equation} Of course one would like to know how this affects gauge choice. To be specific, we would like to know if this new vector potential function, $\alpha\nabla\beta$, obeys the Coulomb gauge condition and if not what physically motivated restrictions would lead to it. The divergence of the new scalar potential function is \[ \nabla . {\bf A}_{\it EP}=\nabla.(\alpha\nabla\beta)=\nabla\alpha .\nabla\beta +\alpha\nabla^2\beta.\] We need to consider divergence theorem, to be more explicit we need the Green's identity of the first kind, in order to resolve the right hand side to this equation. On applying divergence theorem to the $A_{\it MP}=\alpha\nabla\beta$, where $\beta$ is twice differentiable then, we obtain;
\begin{eqnarray}\label{eqthis}\iiint_{V}\nabla.(\alpha\nabla\beta)dV &=& \iiint_{V} (\alpha\nabla^2 \beta+\nabla\alpha.\nabla\beta)dV\nonumber\\ &=&\iint_{C}\alpha\nabla\beta.d{\bf S},\end{eqnarray} where $V$ is a typical volume and $C$ the boundary of enclosing $V$. The special sub-case of equation (\ref{eqthis}) is when $\nabla\beta$ is orthogonal to the oriented surface ${\bf S}$. This, together with the fact that the volume $V$ can be made arbitrarily small, leads to $\nabla.(\alpha\nabla\beta)=0$. We note that in general $\nabla\beta.d{\bf S}\ne0.$  Given our orthogonal requirement, we now have a vector potential function that is defined by two scalar functions. 
Clearly the rate at which the magnetic vector potential ${\bf A}_{MP}$ grows will be modified to: 
\begin{eqnarray}
\label{vecpoteq:1}\frac{d{\bold{A}_{\it MP}}}{d{t}}&=&{\bf U}\times\nabla\times{\bf A}_{\it MP}+{\bf A}_{\it EP}+\eta\nabla^2\bold{A}_{\it MP} -\nabla\phi,\nonumber\\\end{eqnarray} where $\eta$ is the coefficient of diffusion and $\phi$ is any general scalar potential function. Taking the curl of equation (\ref{vecpoteq:1}) and using relevant vector identities we get:
\begin{eqnarray}
\label{eq:0}\frac{d{\bold{B}}}{d{t}}&=&\nabla\times(\bold{U}\times\bold{B})+\eta\nabla^2\bold{B}+(\nabla\alpha\times\nabla\beta),
\end{eqnarray} where again ${\bf B}=\nabla\times{\bf A}_{MP}.$ But how does this affects dynamo theory, if at all it does? 
To investigate this, we need to reexamine the induction equation.
\section{Induction equation}
Maxwell's equations in MKS units are 
\begin{eqnarray}
\frac{\partial {\bf B}}{\partial{t}}&=&-\nabla\times{\bf E}, ~~~~\nabla .{\bf B}=0\\
\epsilon_{0}\mu_{0}\frac{\partial {\bf E}}{\partial{t}}&=&\nabla\times{\bf B}-\mu_{0}{\bf J}, ~~~~\nabla .{\bf E}=4\rho/\epsilon_{0},
\end{eqnarray} where ${\bf E}$ is the electric field, ${\bf J}$ is the current density, $\mu_{0}$ is the magnetic permeability such that the permittivity of free space $\epsilon_{0}=1/c^2\mu_{0}.$ 

In the present context, a further ingredient necessary for establishing the primary equation is Ohm's law. Assuming a plasma fluid, the standard form of this law has the structure
\begin{eqnarray} E+{\bf U}\times{\bf B}=\eta{\bf J}+\frac{1}{ne}{\bf J}\times{\bf B}-\frac{1}{ne}\nabla P,\end{eqnarray} where we have neglected the time variation of the current density $\partial {\bf J}/\partial{t}$. $\eta$ is the magnetic diffusivity, and expresses the relationship between electric field, moving charges of a given density and a magnetic flux. $n$ is the electron number density and $e$ is the charge. Now consider the case where the Hall's term (${\bf J}\times{\bf B}/ne$) may be neglected and where the gradient of pressure term may be replaced by the product of a scalar potential and the gradient of a different scalar potential; an Euler potential term (we will return to this correspondence later in the article). Suffice it to say, for now, that the charges experience the effect of a vector potential $\alpha\nabla\beta$ which lead to a modified Ohm's law of the form $E+{\bf U}\times{\bf B}+\alpha\nabla\beta=\eta{\bf J}$. The standard approach is to choose a time scale in which the displacement current can be neglected. This implies that the time derivative of the electric field is set to zero leading to $\nabla\times{\bf B}={\bf J}$ and since $E=-{\bf U}\times{\bf B}-\alpha\nabla\beta+\eta{\bf J}$ from our modified Ohm's law, it is easy to see that the evolution equation for the magnetic flux is given by
\begin{eqnarray}
\label{flux}\frac{d{\bold{B}}}{d{t}}&=&\nabla\times(\bold{U}\times\bold{B}+\alpha\nabla\beta-\eta\nabla\times\bold{B}.)
\end{eqnarray} The ${\bf U}\times{\bf B}$ is the usual {\it induction} term. It would appear that the term $\alpha\nabla\beta$ plays a role complementary to the inductive term. Traditionally, the velocity term stretches the field thereby enhancing its strength. In contrast, the Euler term seems to be {\it direction focussing} through the cross-product of the gradients of the two scalars. Could this form of constructive inference of two surfaces lead to an enhancement of the field? Indeed it is the role that this term plays that we seek to understand. The useful form of equation (\ref{flux}) is \begin{eqnarray}
\label{flux2}\frac{d{\bold{B}}}{d{t}}&=&\nabla\times(\bold{U}\times\bold{B})+\eta\nabla^2\bold{B}+(\nabla\alpha\times\nabla\beta).\end{eqnarray} It is worth pointing out recent developments related to Euler potentials. In \cite{Axelp}, $\nabla\times {\bf A}_{MP}$ is compared to $\nabla\times {\bf A}_{EP}$ where artificial viscosity is used in simulations involving the latter. This follows the extensive application of $A_{EP}$ in the SPH approach \cite{Pric3, Pric4} to structure formation. It is found in \cite{Axelp} that the $\nabla\times{\bf A}_{EP}$ compares with $\nabla\times {\bf A}_{MP}$ when $\eta=0$ and that the introduction of $\eta\ne0$ leads to discrepancies. The conclusion is that ${\bf A}_{EP}$ and ${\bf A}_{MP}$ cannot be used interchangeably, in particular that the the growth of $\nabla\times {\bf A}_{EP}$ cannot be ascribed to a dynamo effect. Be that as it may, equation (\ref{flux}) suggests that ${\bf A}_{EP}$ may play a role in the amplification of $\nabla\times {\bf A}_{MP}$, which is not in contradiction to \cite{Axelp}. But what other role might this potential play?

\section{Pressure, density and Euler potentials}
Returning to equation (\ref{eqthis}) and the ensuing discussion, $\alpha\nabla\beta$ looks structurally like $\nabla P/\rho$ suggesting the correspondence:  
\begin{eqnarray}
\label{abov}\nabla\alpha\times\nabla\beta\Leftrightarrow\frac{\nabla P\times\nabla \rho}{\rho^2}. 
\end{eqnarray} The implied correspondence in equation (\ref{abov}) calls for further analysis. It is conceivable that in an idealized fluid flow, density and pressure may exhibit properties similar to Euler potentials and thus raising the prospect of more complex interactions between fluid and magnetic field lines, over and above the back reaction via Lorentz force. But equation (\ref{flux}) and the correspondence equation (\ref{abov}) show that if the position of the two gradients of scalar are switched, this term acts as diffusive term. This may partly account for the results obtained in \cite{Axel2010}.We have only dealt with a simplified case of fluid flow where the Lorentz force is switched off because our interest is in examining how Euler potentials affect the threshold for dynamo action and not assessing the long term contribution to the evolution of magnetic fields.

\section{Discussion and Conclusion}
In this rather inquisitively oriented article we have delved into the subject of the dynamo theory. Our interest was peaked by the need to reexamine the basis for dynamo theory in a flow where Euler potentials are present. We have done four things (1) In deriving the induction equation, we find that the following form of Ohm's law, ${\bf E}=-{\bf U}\times {\bf B}+\eta{\bf J}-\alpha\nabla\beta$ is inevitable. The $\alpha\nabla\beta$ acts as a source for electric currents  and needs further investigations. (2) We have also shown that density and pressure variables are nature candidates for Euler potentials for specialized flows in which $\nabla P$ and $\nabla\rho$ are conserved but not parallel. In fact, the $\nabla p/\rho$ is itself a vector potential which sources magnetic flux. The curl of this vector potential is the basis for Biermann battery term in magneto-genesis. (3) In the appendix, we show that Backus theorem for magnetic induction holds even when Euler potentials are taken into account.(4) We find that indeed the strength of strain tensor required for the dynamo action to kick-in would be lower when the effect of such potentials is taken into account. Although this article speaks of the dynamo theory, one can certainly examine the different kinds of dynamos in flows that admit such potentials. In Simulation comparing the {\bf EP} method and the {\bf A } approaches were performed in \cite{Axelp} , where it was found that two potentials exhibit different growth patterns when diffusivity was included. In our case we argue that the two potentials are not interchangeable but complementary. 

\appendix
\section{Energetics and the dynamo action}
The concept of a dynamo is that it is a mechanism that allows for kinetic energy in a system to be converted into magnetic energy \cite{Axel}. One presupposes that such a system is isolated and the growth of magnetic energy is attributed to the mechanism and not to an external contributor. The magnetic energy of such a system is given by
\begin{eqnarray}\label{energ1}
\mathcal{E}_{m}=\iiint\frac{1}{2\mu}|{\bf B}|^2dV.
\end{eqnarray} Our interest is in how magnetic energy changes, more specifically increases, with respect time. In order to estimate the change in magnetic energy, We take the time derivative of equation (\ref{energ1}) and express it as follows:
\begin{eqnarray}\label{energ2}
\mu\frac{d\mathcal{E}_{m}}{d t}=\iiint_{V} {\bf B}.\frac{d{\bf B}}{d t}dV,
\end{eqnarray} where the righthand side is an integral over a volume $V$ of finite conductivity. The term $d{{\bf B}}/d t $ may then be eliminated using magnetic induction equation (\ref{flux}) to give 
\begin{eqnarray}
\label{eq:2}\mu\frac{d}{dt}(\iiint_{V} \frac{1}{2}|{\bf B}|^2dV)&=&\iiint_{V}{\bf B}{.}(\nabla\times({\bf U}\times{\bf B}))dV\nonumber\\&~~~+&\eta\iiint_{V}{\bf B}.(\nabla^2\bold{B})dV\nonumber\\&+&\zeta\iiint_{V} {\bf B}.{\bf B} dV,
\end{eqnarray}
note that one can expanded the first term on the right hand side of equation (\ref{flux}) into the constituent terms; the stretching, the advection and the compression and have used the assumption that the magnetic flux is homogeneous. We also used definition (\ref{zetadef}) in our substitution to obtain the last term in equation (\ref{flux}). The magnitude of the terms of on the right hand side of this equation has to be positive. Several treatments appear in literature  of ways of assessing this. We use Backus approach \cite{Back1}. Following this formulation, use vector manipulations to that ${\bf B}.(\nabla^2\bold{B})$ is equivalent to $|\nabla\times{\bf B}|^2$. In order for a dynamo to work, the diffusive term must be significantly less the the amplifying term. How significantly less should diffusion be? this can be answered by determining a suitable scale for making quantitative comparison. To this end we define the following parameters:
 \begin{eqnarray}
 \label{eq:4} m(t)&=&max(d_{i}u_{j}+d_{j}u_{i})\\
  s_{d}&=&min\frac{\eta\int_{V}{\bf B}.(\nabla^2\bold{B})dV}{\frac{d}{dt}\int_{V'}|{\bf B}|^2 dV}\\
  f_{d}&=&\frac{\zeta\int_{V}{\bf B}.{\bf B}dV}{\frac{d}{dt}\int_{V'}|{\bf B}|^2 dV}, 
  \end{eqnarray} where $m(t)$ is the the maximum of the rate of strain tensor. This leads to the modified inequality 
  \begin{eqnarray}
\label{eq:5}\frac{1}{2}\frac{d}{dt}\int_{V+V'}|{\bf B}|^2 dV\leq (m(t)+f_{d}-s_{d})\frac{d}{dt}\int_{V+V'}|{\bf B}|^2 dV. \nonumber\\
\end{eqnarray} $V+V'$ indicates an integral over the entire space, whereas $V'$ only covers a part of $V$ and arises from the partial
integration of the term $\nabla\times(\eta\nabla\times{\bf B}).$ Dynamo action occurs as long as the net effect of the {\it stretching} $m(t)$ and the {\it focussing} $f_{d}$ is greater than the dissipative term $s_{d}.$ i.e. $m(t)+ f_{d} > s_{d}$. The implication is that the maximum value of the strain tensor need not be as high as previously thought if focussing is taken into account. 

\section{Acknowledgement}
The author acknowledges support from the University of Cape Town's URC and NGP funds.

\end{document}